\documentclass[draft]{agujournal2019}
\draftfalse
\journalname{Geophysical Research Letters}

\usepackage[hidelinks,final]{hyperref}

\makeatletter\def\NAT@sort{\z@}\makeatother

\begin{document}

\title{Solving the mystery of the electron radiation belt at Uranus: Leveraging Knowledge of Earth's Radiation Belts in a re-examination of Voyager 2 Observations}

\authors{R. C. Allen\affil{1}, S. K. Vines\affil{1}, and G. C. Ho\affil{1}}

\affiliation{1}{Southwest Research Institute, San Antonio, TX.}

\correspondingauthor{Robert C. Allen}{Robert.Allen@swri.org}

\begin{keypoints}
\item A comparison of the Voyager 2 Uranus flyby and similar energetic electron event at Earth is presented
\item Solar wind transients and interactions are potentially important at Uranus for driving magnetospheric dynamics, like exciting chorus waves
\item Both events demonstrate chorus wave electron acceleration, resulting in higher intensity electron belts than ion belts
\end{keypoints}

\begin{abstract}
Voyager 2 provided the only direct measurement of the radiation environment at Uranus and established the well-accepted characterization of a system with a weaker ion radiation belt and surprisingly intense electron radiation belt. Recent re-analysis of the flyby, however, suggests that these observations were not taken during normal solar wind conditions, and instead occurred while a large solar wind transient, a corotating interaction region, was passing over the system. With this context, this study compares the Voyager 2 observations to a similar event observed at Earth. This comparative approach, along with an expanded contextual interpretation of the Voyager 2 flyby, suggests that solar wind-magnetospheric interactions at Uranus may have driven enhanced lower band chorus wave emissions capable of accelerating electrons to relativistic energies. This motivates new, more expansive questions to be explored at Uranus and highlights the need for a Uranus orbiter mission.
\end{abstract}

\section*{Plain Language Summary}
Voyager 2 provided the only direct measurement of the radiation environment at Uranus. These findings established the well-accepted characterization of a system with a weaker ion radiation belt and surprisingly intense electron radiation belt. However, these observations may not have occurred during normal conditions, as noted in recent studies. With this context, this study compares the Voyager 2 observations to a similar event observed at Earth. This comparative approach, along with a recent reinterpretation of the Voyager 2 flyby, suggests that interaction between the solar wind and magnetosphere of Uranus may have resulted in enhanced electromagnetic waves capable of accelerating electrons to relativistic energies. This motivates new, more expansive questions to be explored at Uranus and highlights the need for a Uranus orbiter mission.

\section{Introduction}

Uranus has a unique and poorly explored magnetosphere. The inclination of the Uranian equatorial plane with respect to the northward orbital plane is 97.7$^{\circ}$, resulting in times when the rotation axis is oriented very near the planet-Sun line, as well as times when it would be roughly tangential to this line, causing unique seasonal variations in solar wind-magnetospheric dynamics (e.g., \citealp{Connerney1993b}). Additionally, the dipole term of the magnetic moment of Uranus is displaced from, and inclined by 60$^{\circ}$ from, the rotation axis \citep{Ness1986}. The interpretation of the magnetic moment displacement is due to higher order moments of the system, such as a large quadrupole and octupole terms, leading to a complex field geometry \citep{Connerney1987,Stanley2006}. This complex system has important implications for energetic particle acceleration, loss, transport, and trapping, as well as plasma convection (e.g., \citealp{Selesnick1988}; \citealp{Gao1998}; \citealp{Acevski2024}).

The Voyager 2 flyby of Uranus in 1986 provided the only direct measurements of the system to date. During this flyby, Uranus was configured such that the rotation axis was roughly along the planet-Sun line. Despite the complexity of the magnetic field, both ion and electron radiation belts were observed by Voyager 2, which reached $\sim$4 R\textsubscript{U} (Uranus radius, with 1 R\textsubscript{U} = 25,559 km) from Uranus at closest approach \citep{Krimigis1986}, allowing a momentary characterization of their content. While the ion radiation belt intensity was below expectations, the electron belt intensities were significantly higher than anticipated, reaching the Kennel-Petschek limit for the system (e.g., \citealp{Mauk2010}, \citealp{Mauk2014}). The Kennel-Petschek limit is the point at which runaway wave growth would subsequently scatter particles into the loss-cone, thereby limiting the maximum energetic particle intensity that could stably exist (see \citealp{Kennel1966}; \citealp{Acevski2024}). As result of this discrepancy between the ion and electron intensities compared to their Kennel-Petschek limits, studies have posed a seeming mystery regarding how the Uranian system could support such intense electron belt with a meager ion belt (e.g., \citealp{Krimigis1986,Kurth1991,Mauk2010,Mauk2014,Rymer2019,Kollmann2020,Paty2020}).

While signatures suggestive of substorm-type injections were observed, which could seed particles to the inner magnetosphere (e.g., \citealp{Behannon1987,Mauk1987}), low energy plasma observations suggested very low densities during the time of the flyby \citep{McNutt1987}. However, it was recently noted that a co-rotating interaction region (CIR), a large-scale solar wind structures that form at the boundary between high-speed streams from coronal holes and preceding slow solar wind streams (see the review by \citealp{Richardson2018}), was passing over Uranus at the time of the Voyager 2 flyby \citep{Jasinski2024}. As noted in \citet{Jasinski2024}, the passage of the CIR may have caused plasma in the inner magnetosphere to be vacated by the time of Voyager 2's passage through the system. Additionally, complex convection patterns expected for a higher-order magnetic moment system, together with the unique configuration of rotation axis with respect to the dipole tilt, suggests that Voyager 2 may have traversed sparsely populated lobe-like field lines rather than a plasma disk. Either circumstance could have resulted in non-representative observations of the nominal state of the system, or of conditions just preceding the flyby. Regardless of the cause for the thermal population measurements, the presence of a CIR, then, could provide a new explanation for the observed electron belt intensities and motivate deeper scientific questions regarding energetic particles in the magnetosphere of Uranus.

In the following decades since the Voyager 2 Uranus flyby, there has been a tremendous amount of work refining understanding of radiation belt physics. One such advancement was an understanding of how whistler-mode chorus waves can act to rapidly accelerate electrons to relativistic energies (e.g., \citealp{Thorne2013,Reeves2013,Li2019,Miyoshi2021}) and how this process can be aided by pre-conditioning of seed populations (e.g., \citealp{Baker1986}; \citeyear{Baker1998}; \citealp{Boyd2014}). These processes at Earth are known to primarily occur during geomagnetically disturbed times due to solar wind-magnetospheric interactions, such as the impact of CIRs on the Earth's magnetosphere-ionosphere system (e.g., \citealt{Allen2023}). Previous studies have demonstrated that Alfv\'enic fluctuations in the high-speed streams of CIRs during southward IMF result in persistent substorm activity \citep{Tsurutani1987}, which enhance chorus wave excitation by sourcing and modulating seed electrons that drive efficient electron radiation belt acceleration (e.g., \citealp{Miyoshi2003}, \citeyear{Miyoshi2013}; \citealp{Miyoshi2008}). A comparison of the Voyager 2 Uranus flyby to Earth observations from the Van Allen Probes (VAP) mission may then allow for further insight into the dynamics operating within the Uranian system.

\section{Solar Wind and Planetary Magnetospheric Datasets}

Voyager 2 launched in 1977 to take measurements of the magnetospheres of Jupiter, Saturn, Uranus, and Neptune as well as to explore the outer heliosphere and interplanetary medium \citep{Kohlhase1977}. This study investigates the energetic particle content at Uranus using measurements from the Low Energy Charged Particle (LECP; \citealp{Krimigis1977}) experiment, specifically, the ``PL'' ion channels spanning 24 keV to 3.5 MeV and the ``EB'' and ``EG'' electron channels spanning 13 keV to 5 MeV. Additionally, magnetic field observations from the Magnetic field experiment \citep{Behannon1977}, electric field wave-spectra from the Plasma Wave System (PWS; \citealp{Scarf1977}), and plasma moments are from the Plasma Science Experiment (PLS; \citealp{Bridge1977}) are used. The electric field wave-spectra spans 10 Hz to 56.2 kHz with 16 logarithmically space channels.

VAP \citep{Stratton2013} consisted of two identical spacecraft launched in 2012 into Earth orbit to study the Van Allen radiation belts. The Earth-based event in this study (2019 August through September, see Section 3) occurred after VAP-B had deorbited, and so only data from VAP-A is used. Observations from the Relativistic Electron Proton Telescope (REPT; \citealp{Baker2013}, \citeyear{Baker2021}), part of the Energetic Particle, Composition, and Thermal Plasma (ECT) suite \citep{Spence2013} are utilized. For electrons (specifically the 1.8, 4.2, and 7.7 MeV channels), the REPT omni-directional product computed by \citet{Boyd2021} is used, while 21.25 MeV protons are from the spin-averaged product. To better visualize the radiation belts, the REPT observations are binned into 1-hr and 8-hr temporal bins and by L-shell bins 0.25 R\textsubscript{E} wide. To study wave-emissions, magnetic field observations from the Electric and Magnetic Field Instrument Suite and Integrated Science (EMFISIS; \citealp{Kletzing2013}), specifically the tri-axial magnetic search coil (MSC), was used. MSC measures magnetic field fluctuations in the frequency range from about 30 Hz to 12 kHz. This study uses publicly available MSC spectra along the spin-axis direction (i.e., B\textsubscript{W}). Lastly, magnetic field magnitude measurements from the EMFISIS fluxgate magnetometer (MAG; see \citealp{Kletzing2013}) are used. Observations from the OMNI dataset \citep{King2005} are used for upstream solar wind conditions at Earth.

\section{Comparing Plasma Conditions and Magnetospheric Response at Uranus and Earth}

Figure 1 shows the first 13 years of Voyager 2 observations (left, panels a-d) and 13 years of near-Earth observations (right, panels f-i) covering approximately the same portions of two solar cycles (right, panels j-k). The Voyager 2 density and dynamic pressure (P\textsubscript{dyn}) are scaled to 1 au using empirically derived scaling laws, i.e. density and P\textsubscript{dyn} scale as $r^{-1.96}$ \citep{Perrone2019} to allow a more direct comparison. For the solar wind conditions, 27-day averages (i.e., over a Carrington rotation) of the solar wind speeds, densities, and P\textsubscript{dyn} are included (blue line in panels b-d and g-i) to show the long-term averages over the course of the solar cycles. As shown by the vertical lines marking the Voyager 2 planetary flybys, only the Uranus flyby occurred during solar minimum (Figure 1a). Comparing Figure 1b and 1g, the solar wind speed is more variable at 1 au than further out in the heliosphere due to the merging of CIRs at large heliospheric distances (e.g., \citealp{Burlaga2003}). As such, as Voyager 2 reaches larger distances, the variability in solar wind speed and scaled density and P\textsubscript{dyn} decreases.

\begin{figure}[tbp]
\centering
\includegraphics[width=\textwidth]{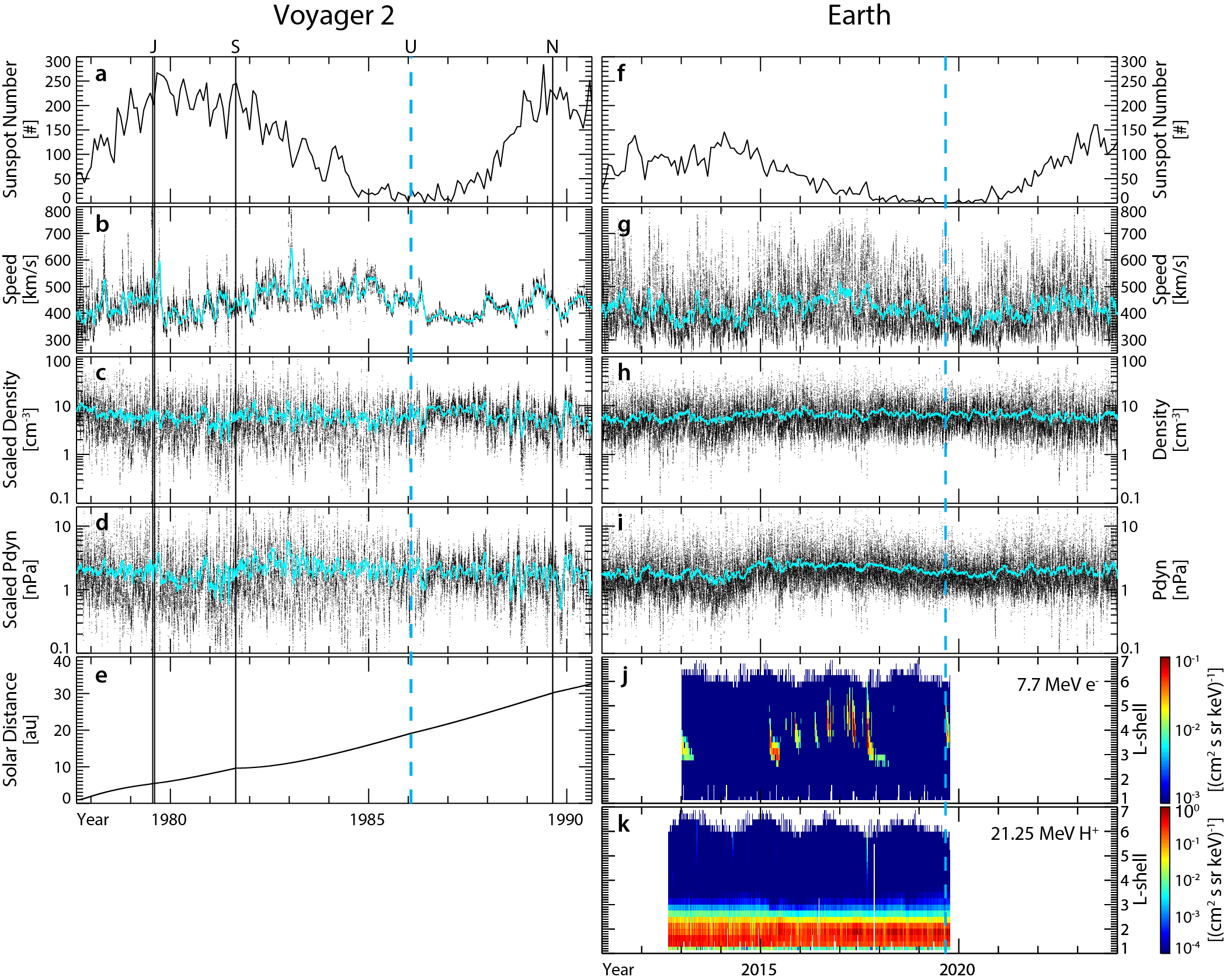}
\caption{(left) Observations from the first 13 years of Voyager 2 operations and (right) 13 years of observations encompassing the VAP mission. (a and f) Monthly sunspot number, (b and g) solar wind speed, (c and h) density, and (d and i) P\textsubscript{dyn}. For comparison, the density and P\textsubscript{dyn} for Voyager 2 are scaled to 1 au values (see text). For context, (e) solar distance of Voyager 2 along with vertical lines marking the flybys of Jupiter, Saturn, Uranus, and Neptune (labeled J, S, U, and N, respectively) are provided. (j) 7.7 MeV electron flux and (k) 21.25 MeV H\textsuperscript{+} flux over the entire VAP mission. The blue dashed vertical line on right panels mark a pronounced 7.7 MeV electron flux enhancement during solar minimum.}
\label{fig:1}
\end{figure}

The flux of 7.7 MeV electrons and 21.25 MeV H\textsuperscript{+} for the entirety of the VAP mission are shown in Figure 1j and 1k, respectively. As noted in other studies, the main H\textsuperscript{+} radiation belt between L-shells of about 1 and 2 is largely stable, even during disturbed times (e.g., \citealp{Xu2019}). However, the electron radiation belt, which can extend between L-shells of about 3 and 6, is highly variable based on magnetospheric activity (e.g., \citealp{Claudepierre2020a}; \citeyear{Claudepierre2020b}; \citealp{Baker2021} and references there in). As shown in Figure 1j, several pronounced events impacted electrons populations up to the 7.7 MeV, one of which occurred during solar minimum (indicated by the dashed blue line).

To further compare conditions during the Voyager 2 Uranus flyby and the solar minimum high-energy electron event at Earth, Figure 2 provides 10 months of unscaled solar wind observations near Uranus (Figure 2a-c) and Earth (Figure 2d-f). Both time periods are characterized by periodic CIRs, typical for solar minimum (e.g., \citealp{Richardson2018,Allen2019,Gibson2023}). While these did not lead to significant variability in the 21.25 MeV radiation belt H\textsuperscript{+} populations (Figure 2i), these CIRs did result in periodic enhancements in the 4.2 MeV radiation belt electrons (Figure 2g). Only the CIR impacting Earth on 2019 August 30 -- September 1 resulted in observable radiation belt electron acceleration up to 7.7 MeV (Figure 2h). A more in-depth investigation of these recurring CIRs and associated heliospheric, magnetospheric, ionospheric, and thermospheric impacts is given in \citet{Allen2023}.

\begin{figure}[tbp]
\centering
\includegraphics[width=0.58\textwidth]{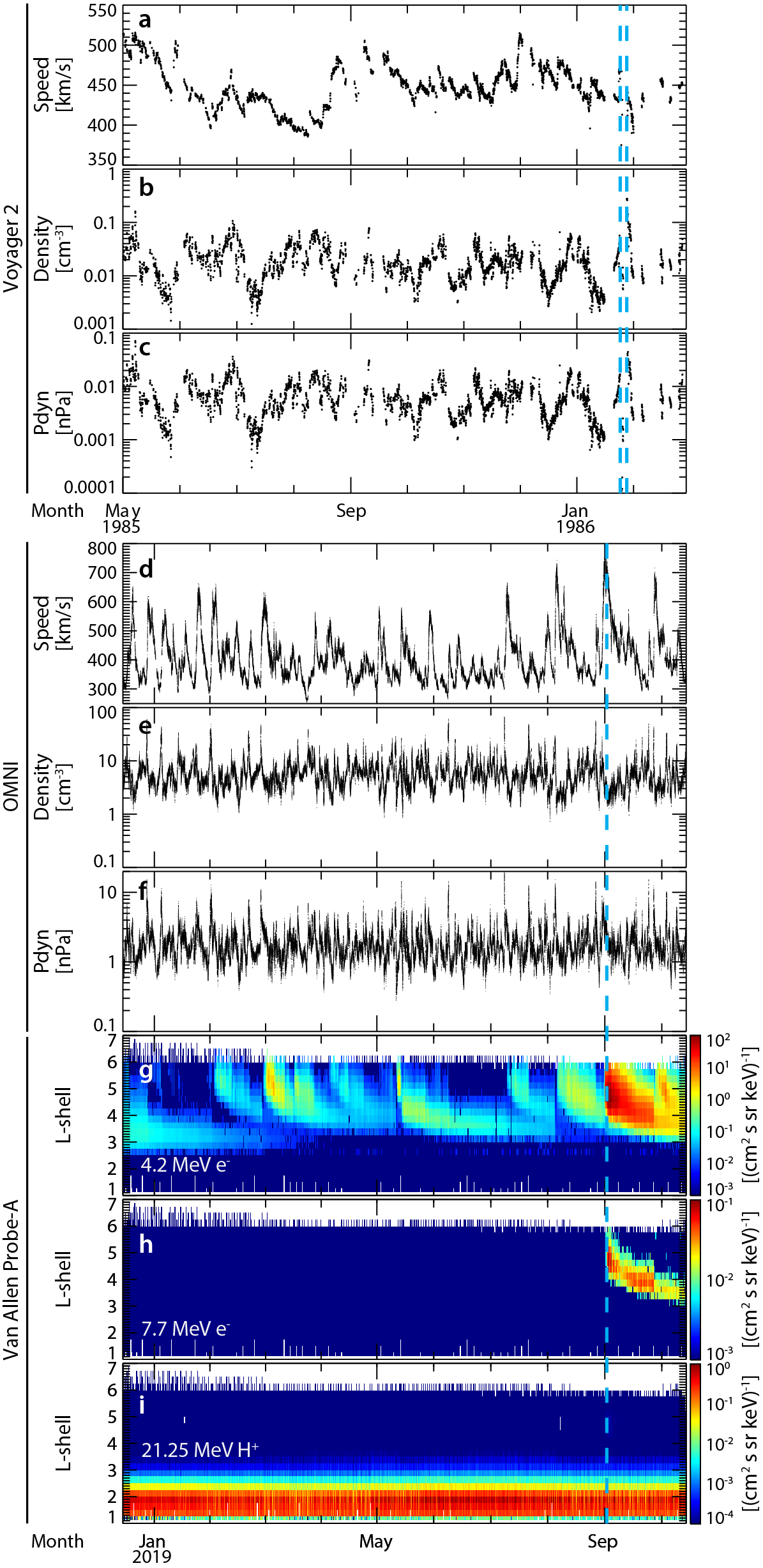}
\caption{Voyager 2 observed solar wind (a) speed, (b) density, and (c) P\textsubscript{dyn} with the crossing through the Uranus system indicated by the blue dashed lines. OMNI solar wind (d) speed, (e) density, and (f) P\textsubscript{dyn}, along with VAP (g) 4.2 MeV and (h) 7.7 MeV electron flux and (i) 21.25 MeV H\textsuperscript{+} flux. The blue dashed line for panels d-i denote the start of the 7.7 MeV electron enhancement.}
\label{fig:2}
\end{figure}

Figure 3 shows two days around the Voyager 2 Uranus flyby. The magnetic field in the U1 Uranus coordinate system during the closest encounter and in Heliographic (HG) coordinates when further away are provided in Figure 3a and 3b, respectively. The magnetic field strength is much stronger near the planet than further away, as expected, with clear magnetotail structures observed in the outbound portion of the flyby (see \citealp{Ness1986}; \citealp{Connerney1987}; \citealp{Behannon1987} for more discussion of the observed magnetic field). The electric field power spectral density from PWS (Figure 3c) shows various wave emissions during the flyby (see \citealp{Gurnett1986}; \citealt{Kurth1991} for detailed descriptions of all waves observed). Due to the large magnetic field near the planet (Figure 3a), the electron cyclotron frequency (\textit{f}\textsubscript{ce}) is within the PWS frequency range near the planet (solid white line on Figure 3c, with 0.5\textit{f}\textsubscript{ce} and 0.1\textit{f}\textsubscript{ce} also marked by the white dashed and magenta dotted lines, respectively). Notably, intense lower band chorus waves are detected during the flyby, as seen by the narrow-band emission between 0.1\textit{f}\textsubscript{ce} and 0.5\textit{f}\textsubscript{ce} (Figure 3c). As noted in \citet{Kurth1991}, these emissions were the strongest observed by PWS at any planet.

\begin{figure}[tbp]
\centering
\includegraphics[width=\textwidth]{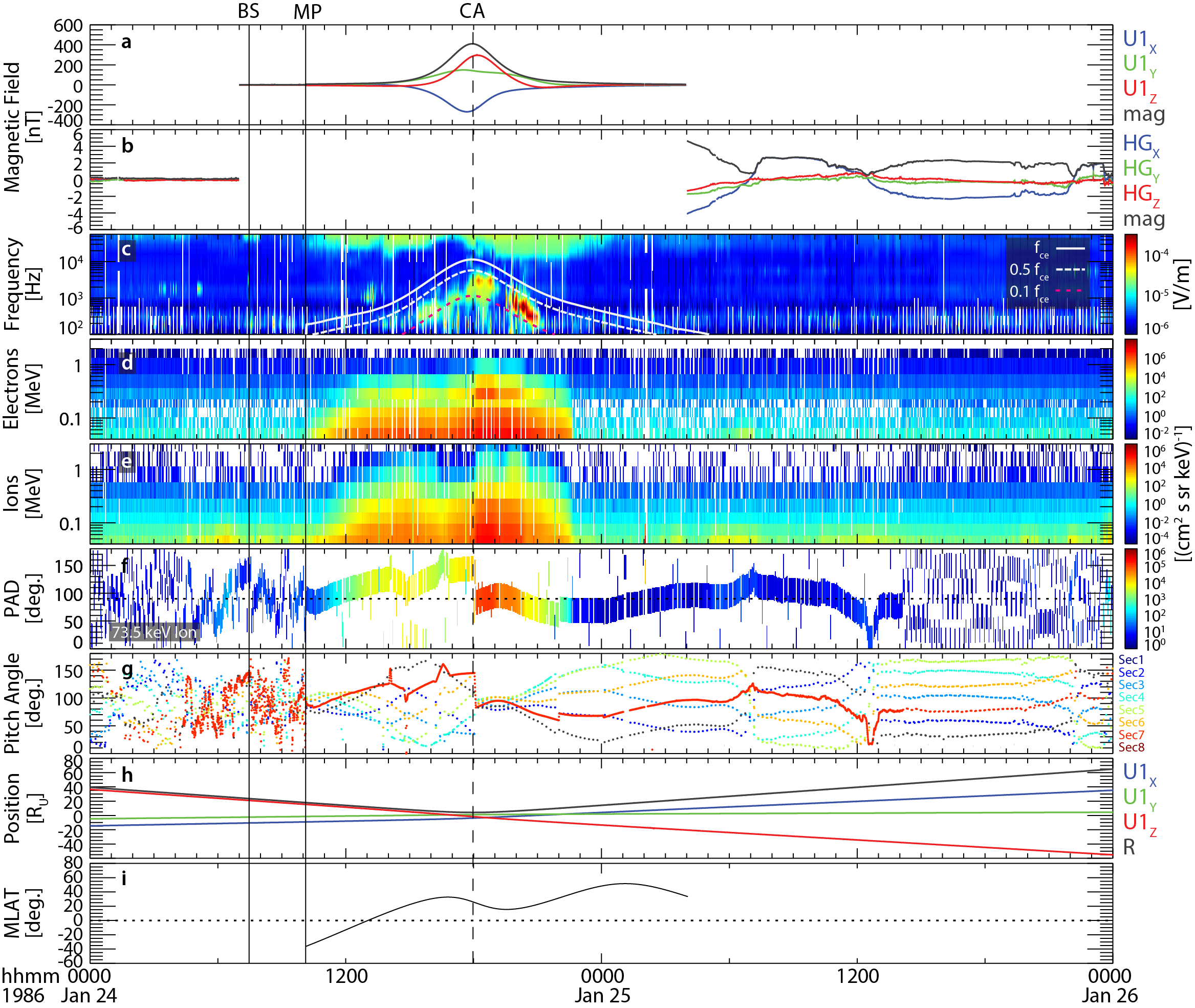}
\caption{Voyager 2 magnetic field measurements in (a) U1 coordinates and (b) HG coordinates, (c) electric field power spectral density, (d) electron and (e) ion flux spectrograms, (f) pitch angle distributions of 73.5 keV ions, (g) centroid pitch angles of LECP observations, (h) location in U1 coordinates, and (i) MLAT assuming a dipolar field. Vertical lines denote the bowshock crossing (BS), magnetopause crossing (MP) and closest approach (CA). Lines within panel c denote the electron cyclotron frequency (\textit{f}\textsubscript{ce}), 0.5\textit{f}\textsubscript{ce}, and 0.1\textit{f}\textsubscript{ce}.}
\label{fig:3}
\end{figure}

From the formulation of \citet{Summers2007}, and electron densities for times of the flyby when the cold electron population is observable (see \citealp{Sittler1987,Belcher1991}), the upper bound for the minimum resonance energy for chorus emissions near 0.5\textit{f}\textsubscript{ce} is in the tens of keV with an average of $\sim$75 keV. This suggests that the observed chorus waves would be resonant with most if not all electrons observed by LECP, however more detailed analysis to estimate the energy gain through deriving the full diffusion coefficients for the Uranian system is required. Additionally, the broadening of the radio wave emissions above $\sim$10\textsuperscript{4} kHz from the inbound to outbound portions of the flyby may be indicative of substorm activity, also suggested by in situ observation (e.g., \citealp{Behannon1987}), similar to broadening of auroral kilometric radiation emission at Earth associated with substorm activity (e.g., \citealp{Morioka2008}), which may act to supply seed populations to the inner magnetosphere.

The energetic electron and total ion populations (Figure 3d and 3e, respectively) are enhanced near Uranus, suggesting at least quasi-stable radiation belts. While the observed fluxes are higher on the outbound pass of the belts (i.e., after the closest approach), this period also corresponds to times when the LECP look direction was closer to 90$^{\circ}$ in pitch angle for both ions (e.g., Figure 3f) and electrons. As such, it is not possible to determine if the higher flux is solely an impact of the inbound inner magnetospheric portion being at higher magnetic latitude (MLAT; see \citealp{Kurth1991} and Figure 3i), differences in local times, or an impact of different pitch angle distribution (PAD) coverage (Figure 3f). During the flyby, the normal operational mode for LECP was modified to primarily observe from Sector 7, with only occasional steps to other sectors (Figure 3g, see also \citealp{Mauk1987}). Beyond this operational change, all sectors of LECP were roughly aligned near $\sim$90$^{\circ}$ pitch angle in the period just after closest approach. These limitations, coupled with the 45$^{\circ}$ field-of-view of LECP, hampers the ability to reliably fit PADs for this flyby for more detailed analysis. For example, this limitation inhibits the ability to perform data-model comparisons for particle source/loss processes (e.g., \citealp{Glauert2024}), comparisons to known wave-resonance impacts to PADs (e.g., \citealp{Thorne2013}; \citealp{Xiao2015}), and comparisons to statistical PADs at other planetary bodies (e.g., \citealp{Clark2014,Smirnov2022}). However, the times of highest particle flux do appear to correspond near times of intense chorus wave power (Figure 3c-e).

As a point of comparison, Figure 4 presents observations taken during the solar minimum 7.7 MeV electron enhancement at Earth, noted in Figure 2. The orbit of VAP is highly elliptic, allowing it to ``scan'' through L-shells of $\sim$1 to 6. As a result, the magnetic field magnitude varies over a large range between perigee and apogee (Figure 4a). However, \textit{f}\textsubscript{ce} is within the MSC frequency range for most of the orbit (Figure 4b). Lower band chorus waves, seen as the narrow-band emissions between 0.1\textit{f}\textsubscript{ce} and 0.5\textit{f}\textsubscript{ce}, were triggered with the arrival of the CIR spanning 2019 August 30 -- September 1 (Figures 4a and 4b). Corresponding to this, electrons are continually accelerated to progressively higher energies (Figures 4c-e), consistent with local chorus wave-driven acceleration (e.g., \citealp{Horne2005,Thorne2013}), with little change in the 21.25 MeV H\textsuperscript{+} population (Figure 4f). As noted in \citet{Allen2023} and \citet{Hudson2021}, this specific CIR was likely so efficient at accelerating electrons due to pre-conditioning of the magnetospheric system. Under this rationale, previous convection and substorm activity (due to preceding CIRs in 2019 August) could have populated an inner magnetospheric seed populations (e.g., \citealp{Baker1986}; \citeyear{Baker1998}; \citealp{Boyd2014}) which then enabled enhanced acceleration through stochastic processing of the seed population with chorus emissions (e.g., \citealp{Reeves2013}; \citealt{Thorne2013}; \citealp{Li2019}). At Earth, a combination of local chorus-driven acceleration and radial diffusion, such as diffusion resulting from ultralow frequency (ULF) waves which have a higher occurrence during high-speed streams (e.g., \citealp{Regi2015}, \citealp{Teramoto2019}), can also operate simultaneously and contribute differently depending on solar wind and magnetospheric conditions \citep{Takahashi2021}. Additionally, as noted in \citet{Daggitt2024}, in a system leading to strong pitch angle diffusion, the loss due to pitch angle scattering by intense chorus wave may be more limited than the extent of acceleration due to diffusive wave-particle interactions at MeV energies. The combined result is a higher intensity electron radiation belt with nominal intensity H\textsuperscript{+} belt, similar to during the Voyager 2 Uranus flyby.

\begin{figure}[tbp]
\centering
\includegraphics[width=\textwidth]{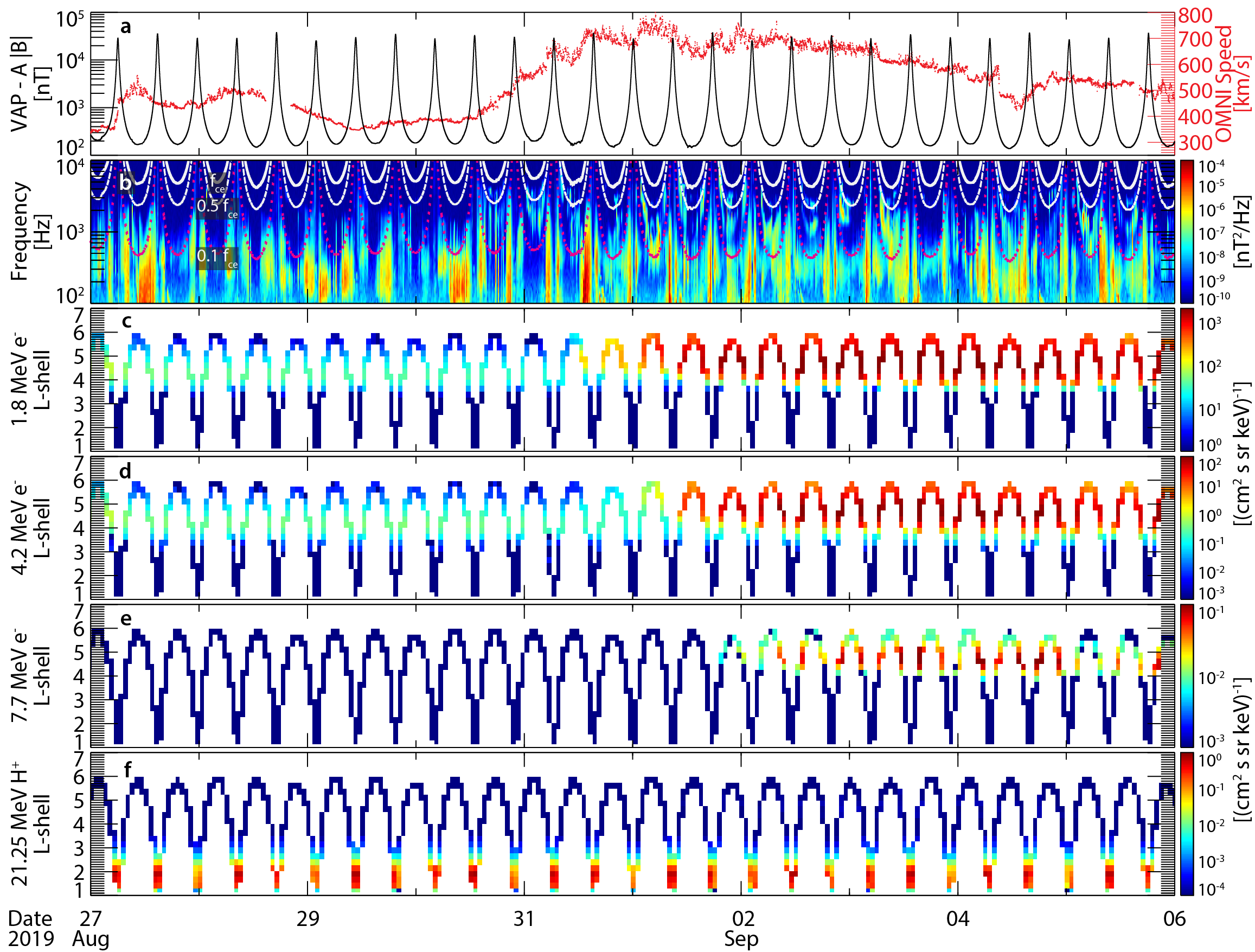}
\caption{(a) Solar wind speed (red) and VAP-A magnetic field magnitude (black), (b) magnetic field power spectral density, and flux of (c) 1.8 MeV, (d) 4.2 MeV, and (e) 7.7 MeV electrons and (f) 21.25 MeV H\textsuperscript{+}. Lines within panel b denote the f\textsubscript{ce}, 0.5f\textsubscript{ce}, and 0.1f\textsubscript{ce}.}
\label{fig:4}
\end{figure}

\section{Conclusions and Open Questions}

The Voyager 2 Uranus flyby and the 7.7 MeV electron event at Earth in 2019 share several similarities. Both occurred during solar minimum when CIRs are more prevalent in the heliosphere (see \citealp{Richardson2018}) and during times when CIRs are seen to be repeatedly passing over the planetary magnetospheres (Figure 2 and depicted in Figure 5). Also, both events are characterized by strong chorus emissions (Figures 3 and 4, shown schematically in Figure 5) concurrent to high fluxes of relativistic electrons, which were potentially driven by convection and substorm activity observed for both Earth (see \citealp{Allen2023}) and Uranus (see \citealp{Behannon1987}), potentially seeding the inner magnetosphere. While the Uranus flyby does not allow for investigations into the time-evolution of the radiation belts, the 2019 event at Earth shows progressive electron acceleration to higher energies, indicative of local acceleration from chorus waves \citep{Horne2005,Thorne2013}. As such, these observations suggest that similar processes related to chorus-driven electron acceleration may be occurring at both Earth and Uranus (Figure 5). This suggests that the high energetic electron flux at Uranus may not be typical and that the cause of this elevated flux can be understood within the current framework used to understand the electron radiation belt dynamics at Earth. Further, the potentially low-density inner magnetosphere at Uranus during this time may also be beneficial for this type of acceleration, as lower plasma densities would make the system more conducive to efficient local momentum diffusion \citep{Horne2005}.

\begin{figure}[tbp]
\centering
\includegraphics[width=\textwidth]{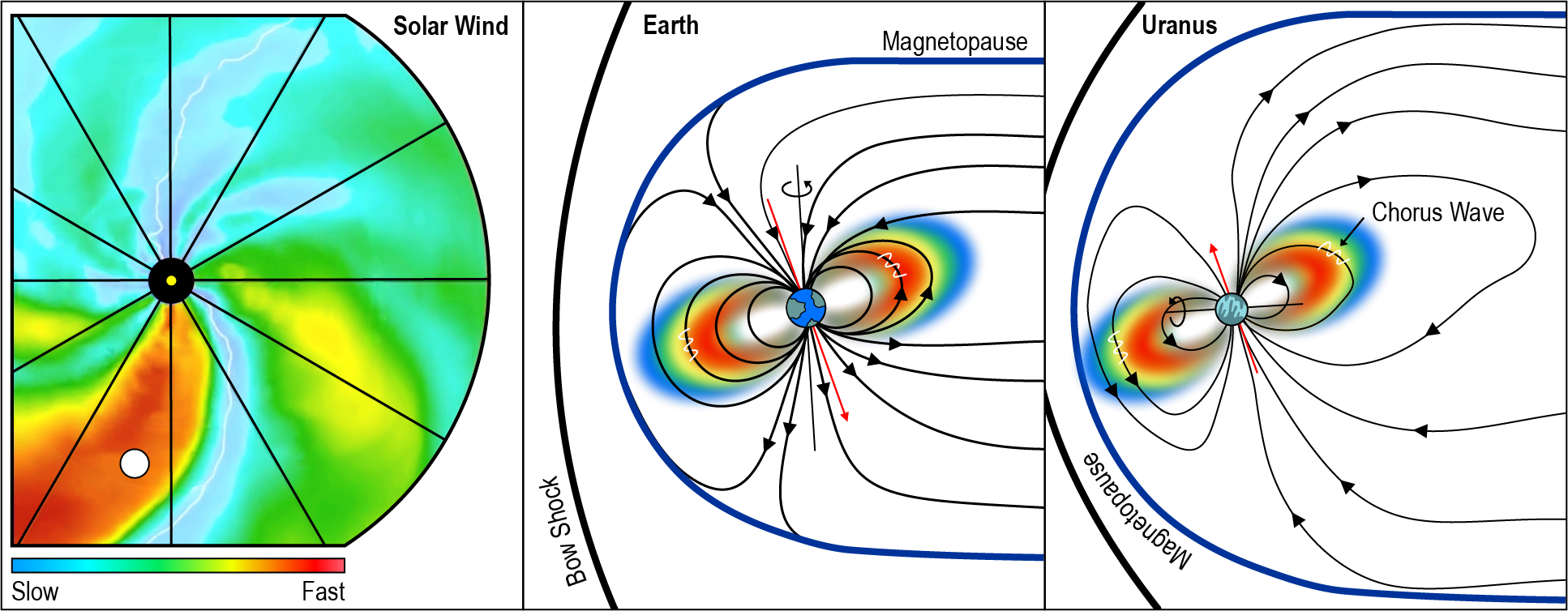}
\caption{Schematic of possible dynamics occurring during enhanced electron radiation belt events at Earth and at Uranus -- not to scale. (left) Both planets are within the high-speed stream of a CIR (representative location shown by the white dot), driving intense Chorus wave emission at (middle) Earth and (right) Uranus that accelerates electrons to relativistic energies. Rotation (black line) and magnetic moment (red arrow) axes are shown for both planets.}
\label{fig:5}
\end{figure}

Lower band chorus wave acceleration arising from a CIR impacting the Uranian magnetosphere may then largely address the reported mystery of the intense electron belt. However, this then motivates a wide range of questions that extend further into understanding the magnetosphere at Uranus, and magnetospheres in general. These questions include:

\begin{itemize}
\item What role does magnetospheric pre-conditioning play at Uranus, and how does it contrast with Earth?
\item To what extent do lower-frequency waves, such as electromagnetic ion cyclotron waves and ULF waves, act to scatter relativistic electrons and heat/accelerate ions at Uranus?
\item How unstable are radiation belts within a tilted magnetosphere and how does this vary through seasons?
\item How does the complexity of the magnetospheric configuration, including the offset between any moon-related neutral tori and the magnetic equator, impact particle diffusion and subsequent wave-particle interactions?
\item What is the contribution of the solar wind, ionosphere, and satellites to sourcing energetic particles in such a system?
\item To what extent do energetic particles contribute to icy moon/ring surface weathering?
\item How do asymmetries and diurnal and seasonal variability in the magnetospheric configuration impact particle precipitation, and how does that contribute to UV and IR auroral emissions?
\end{itemize}

These questions all require a Uranus Orbiter mission to take repeated measurements of the energetic particle population within different regions of the magnetosphere. The addition of refined pitch angle coverage and resolution for such energetic particle instrumentation would aid in the ability to investigate particle acceleration, loss, and trapping within the system. Large energy dynamic range and energy resolution would further advance the ability to robustly explore these processes within this unique magnetospheric system. Finally, compositional determination would be highly beneficial in constraining the sources of particles and gyro-orbit-dependent (i.e., mass or mass per charge dependent) processes on the population. All these aspects of energetic particle dynamics, fundamental to all planetary systems, should be considered in future mission concepts to Uranus. This also motivates additional re-analysis of Voyager 2 planetary flybys given the advances in our knowledge of wave-particle interactions and magnetospheric dynamics at Earth and other systems. Further comparative magnetospheric investigations may then be a useful tool in deepening understanding of fundamental plasma dynamics within widely differing planetary magnetospheres.

\acknowledgments
The authors acknowledge internal support from Southwest Research Institute that enabled this investigation. RCA and SKV also acknowledge support from NASA grant 80NSSC25K7258. The authors thank AJ Galaviz III, Southwest Research Institute, for the creation of Figure 5.

\section*{Conflict of Interest Statement}
The authors have no conflicts of interest to disclose.

\section*{Open Research}
Voyager 2 data are available on the Planetary Data System (PDS), including LECP (\citealp{Mauk1993}; \citeyear{Mauk2012}), MAG (\citealp{Ness1993}; \citealp{Connerney1993a}), and PLS (\citealp{Richardson1993}; \citeyear{Richardson2019}). The Voyager 2 PWS electric field spectrograms, Van Allen Probes EMFISIS data, and OMNI dataset are available via the NASA Space Physics Data Facility at \url{https://spdf.gsfc.nasa.gov/} (specifically, \url{https://cdaweb.gsfc.nasa.gov/pub/data/voyager/voyager2/wave_spectra_pws/spectrum_analyzer_cdf/1986/}, \url{https://cdaweb.gsfc.nasa.gov/pub/data/rbsp/rbspa/l3/emfisis/magnetometer/4sec/gsm/2019/}, \url{https://cdaweb.gsfc.nasa.gov/pub/data/rbsp/rbspa/l2/emfisis/wfr/spectral-matrix-diagonal-merged/2019/}, and \url{https://cdaweb.gsfc.nasa.gov/pub/data/omni/omni_cdaweb/hro2_1min/}). Van Allen Probes ECT data is available at \url{https://rbsp-ect.newmexicoconsortium.org/rbsp_ect.php} (specifically, \url{https://rbsp-ect.newmexicoconsortium.org/data_pub/rbspa/ECT/level3/2019/} and \url{https://rbsp-ect.newmexicoconsortium.org/data_pub/rbspa/rept/level2/sectors/2019/}).

\bibliography{uranus}

\end{document}